# Positive and negative entropy production in thermodynamic systems

José Iraides Belandria

Escuela de Ingeniería Química, Universidad de Los Andes, Mérida, Venezuela

joseiraides @ ula.ve

This article presents a heuristic combination of the local and global formulations of the second law of thermodynamics that suggests the possibility of theoretical existence of thermodynamic processes with positive and negative entropy production. Such processes may exhibit entropy couplings that reveal an unusual behavior from the point of view of conventional thermodynamics.

## 1. Introduction

By the middle of the XX century, Prigogine [1] postulates the local formulation of the second law of thermodynamics expressing that the variation of entropy of a system $dS$ is equal to the entropy flow due to the interactions with the surroundings $d_e S$ plus the internal entropy production $d_i S$ caused by changes inside the system. Thus,

$$dS = d_e S + d_i S \qquad (1)$$

In the case of a closed system $d_e S$ is given by the expression

$$d_e S = dQ / T \qquad (2)$$

where $Q$ represents the heat flow received by the system from the exterior and $T$ is the temperature of the system.

According to Prigogine, the production of internal entropy $d_i S$ in any macroscopic portion of a system is always equal or greater than zero. It is zero when the processes in the system are reversible and it is greater than zero if the system is subjected to irreversible transformations. In other words, $d_i S$ represents the creation or production of internal entropy due to the irreversible processes inside the system like the irreversible flows of heat, mass or moment, among other possibilities.

In addition, Prigogine proposes axiomatically that the destruction or absorption of internal entropy in a part of the system, compensated by an enough production in another region outside of the system are prohibited. Only, this phenomenon can happen, when irreversible processes are coupled inside the same system like the thermo diffusion process or in systems involving coupled simultaneous reactions. For example, if two simultaneous irreversible reactions i, j exhibit entropy couplings inside a system the total production of internal entropy for unit of time $d_iS/dt$ is equal to

$$d_iS/dt = d_iSi/dt + d_iSj/dt \tag{3}$$

Here, $d_iSi/dt$ and $d_iSj/dt$ represent the productions of internal entropy of the processes i, j, respectively. If they are coupled $d_iSi/dt > 0$ and $d_iSj/dt < 0$, or vice versa, but the production of total entropy $d_iS/dt$ should be greater than zero. Prigogine, also, establishes that the entropy coupling should obey an extra-thermodynamic postulate based on the correlations of Onsager which states that the interference coefficients Lij and Lji corresponding to the coupled irreversible processes should be equal and different from zero.

According to Prigogine, his formulation is the only general approach of irreversibility that allows a closer analysis of irreversible transformations because it is focused inside the system where such events take place, in contrast, with the global formulation of the second law supported in classical thermodynamics that outlines the approach based on the irreversibilities happening in the global universe of the process formed by the system and its surroundings.

In this sense, classical thermodynamics [2, 3] formulates the second law in terms of the variation of the total entropy of the universe dSu which should be equal or greater than zero. It is zero when the transformations in the universe are reversible and greater than zero when irreversible events occur. This proposition is known as the global formulation of the second law of thermodynamics and it is expressed by the equation

$$dSu \geq 0 \tag{4}$$

According to classical thermodynamics, the variation of the total entropy of the universe is an additive contribution of the variations of entropy of the different parts that integrate the universe which may be consider constituted by the system and its surroundings. Therefore, the variation of entropy of the universe dSu is equal to the variation of entropy of the system dSs plus the variation of entropy of the surroundings dSa, hence

$$dSu = dSs + dSa \tag{5}$$

**2 . Combination of the local and global formulations of the second law of thermodynamics**

From a creative point of view we can combine the local and the global formulations of the second law of thermodynamics, and extend the underlying ideas of the local formulation to the universe of the process constituted by the system and its surroundings. To simplify, let us suppose that the system and its surroundings are closed and that the universe containing the system and its surroundings is closed and adiabatic, that is to say isolated thermally. Then, applying Ec. (1) to the global universe we obtain

$$dS_u = d_e S_u + d_i S_u \qquad (6)$$

where $d_i S_u$ expresses the total production of internal entropy due to the irreversibilities taking place inside the universe, and $d_e S_u$ describes the entropy flow due to the interactions with the exterior of the universe. In this case, as the described universe is closed and adiabatic, then

$$d_e S_u = 0 \qquad (7)$$

and

$$dS_u = d_i S_u \qquad (8)$$

As it has been explained previously, the global formulation demands that the variation of the total entropy of the universe never is negative, but equal or greater than zero, then, according to Ec. (8), the total production of internal entropy of the universe $d_i S_u$ should be equal or greater than zero.

Now, substituting Ec. (8) in Ec. (6), and applying Ec. (1) to the system and the surroundings, it is obtained

$$d_i S_u = d_e S_s + d_i S_s + d_e S_a + d_i S_a \qquad (9)$$

In this equation the term $d_e S_s$ represents the entropy flow due to the interactions of the system with its surroundings, and $d_e S_a$ expresses the entropy flow associated with the interactions of the surroundings with the system and the rest of the universe. The characters $d_i S_s$ and $d_i S_a$ express the production of internal entropy due to the irreversibilities inside the system and the surroundings, respectively.

As the assumed system is closed, the interactions of the system with the surroundings reduce to the heat flow among both parts, that is to say

$$d_e S_s = dQ_s / T_s \qquad (10)$$

Here $Q_s$ represents the heat flow transferred between the system and the surroundings and $T_s$ is the temperature of the system.

Similarly, as the surroundings are closed, their interactions with the exterior reduce to the heat flow with the system, only, because the universe containing the system and their surroundings is closed and isolated thermally. Under this consideration

$$d_e S_a = dQ_a / T_a \qquad (11)$$

where Qa is the flow of heat between the surroundings and the system and Ta is the temperature of the surroundings.

Then, by combining Ecs. (9), (10), (11) we get

$$d_i Su = (dQs / Ts + dQa / Ta) + d_i Ss + d_i Sa \qquad (12)$$

Evidently, the term (d Qs / Ts + dQa / Ta) represents the production of internal entropy inside the universe $d_i Sm$ due to the irreversible flow of heat between the system and the surroundings. But, according to the first law of thermodynamics $dQa = - dQs$, then

$$d_i Sm = dQs / Ts + dQa / Ta = dQs (1 / Ts - 1 / Ta) \qquad (13)$$

As a reference, Prigogine [1] deduces an expression similar to this equation for the entropy production due to the irreversible flow of heat among two phases maintained at different temperatures.

By combining Ecs. (12) and (13), we find

$$d_i Su = d_i Sm + d_i Ss + d_i Sa \qquad (14)$$

This equation suggests that in the case of a closed and adiabatic universe, the production of entropy of the universe depends on the additive contributions of the productions of internal entropy associated with the system, the surroundings and the heat flow between the system and the surroundings.

As we can perceive Ec. (14) results from a combination of the local and global formulations of the second law of thermodynamics and expresses the global formulation in terms of the entropy production in the different regions that integrate the universe and, in this sense, it is an extension of the local formulation that embraces the whole universe including the system and its surroundings. Therefore, we can propose, with a wide vision, the possibility of existence of internal entropy couplings to the scale of the universe involving interactions between the system and its surroundings. This would be an extension of the internal entropy couplings that happen to the scale of the system in the local formulation of Prigogine. Under this consideration, the destruction or absorption of internal entropy in a region of the universe like the system, compensated by an enough production in another region of the universe, the surroundings, is possible.

In this sense, we can argue that the possibility of existence of such processes requires, only, that the global formulation of the second law of thermodynamics holds, that is to say that the algebraic sum of the terms of Ec. (14) must be equal or greater than zero, independent of the sign, positive or negative, that each term may have. If these processes could happen, they would present an unusual behavior as it is described in the following section.

Another implication of Ec. (14) appears when $dS_u = d_iS_u = 0$. In these circumstances two cases may exist. In the first case $dS_u = d_iS_u = 0$ when the production of internal entropy in all places of the universe is zero, that is to say, $d_iS_m = d_iS_s = d_iS_a = 0$. This solution of Ec. (14) coincides with the classic demand of reversibility in all the events of the universe. The second case, on the other hand, is when some of the terms corresponding to the production of internal entropy are positive and other negatives, but the algebraic sum of them is equal to zero. As an example $d_iS_m > 0$, $d_iS_s < 0$ and $d_iS_a > 0$, but $d_iS_m + d_iS_s + d_iS_a = d_iS_u = dS_u = 0$. This case is not predicted by classical thermodynamics. However, the combination of the local and global formulations of the second law of thermodynamics suggests this possibility.

**3. Model of a process with internal entropy coupling between the system and its surroundings.**

Let us consider the process schematized in Fig. 1 in which two tanks A and B are separated by a good heat conducting metallic partition covered initially by an adiabatic film. Each tank contains 1 mol of a monoatomic ideal gas and the initial conditions in tank A are 4 atm and 500 K and in tank B are 4 atm and 800K. Both tanks, including the piston, are externally covered by an adiabatic wall. To simplify the analysis it is assumed that the heat capacities and the mass of the walls of the tanks and of the metallic partition are negligible.

To begin the process the adiabatic film is removed and the heat flows, irreversibly, from the tank B toward the tank A, due to the difference of temperatures among the contents of the tanks. During the process tank B stays at constant volume, and the transferred heat is used to carry out in tank A an isothermal expansion at 500K. The process concludes when thermal equilibrium between the tanks is reached which happens when the temperature in tank B is 500K.

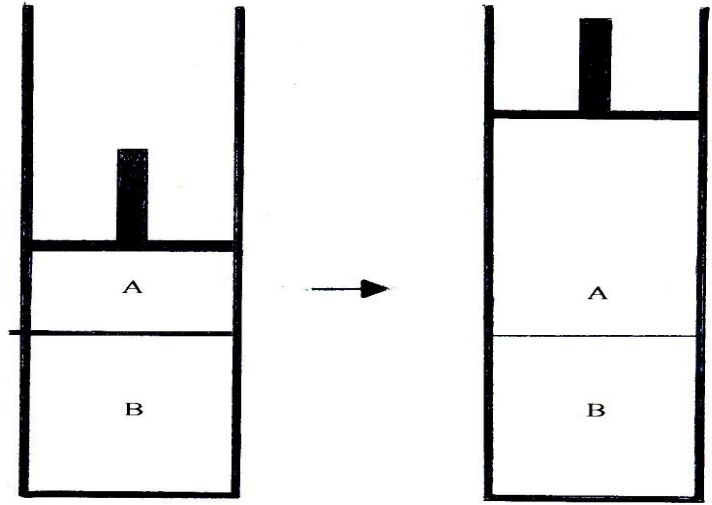

**Fig. 1. Process with internal entropy coupling**

**4. Discussion**

To start, we will apply the global formulation of the second of thermodynamics to the universe of the previous process which is an adiabatic closed universe. By inspection, the universe is constituted by tank A, tank B and the external region of both tanks. Since tanks A and B are closed systems, and their external walls adjacent with the rest of the universe are isolated thermally, then, according to Ec. (5), the variation of the total entropy of the universe dSu is

$$dSu = dS_A + dS_B \qquad (15)$$

where $dS_A$ and $dS_B$ denote the variation of entropy of the ideal gas contained in tanks A and B, respectively. Then, by integrating Ec. (15) we obtain the total entropy change of the universe $\Delta Su$ for the specified change of state

$$\Delta Su = - R \, Ln \, (P_{2A} / P_{1A}) + Cv \, Ln \, (T_{2B} / T_{1B}) \qquad (16)$$

Here R, Cv, $P_{1A}$, $P_{2A}$, $T_{1B}$ and $T_{2B}$ are the ideal gas constant, the heat capacity at constant volume, the initial pressure of tank A, the final pressure of tank A, the initial temperature of tank B and the final temperature of tank B, respectively.

As we can see, the final pressure $P_{2A}$ is not defined previously, but we may know the range of permissible values allowed by the global formulation of the second law of thermodynamics, that is to say, those for which $dSu \geq 0$. Then, according to Ec. (16)

$$(P_{2A} / P_{1A}) \leq (T_{2B} / T_{1B})^{Cv/R} \tag{17}$$

Therefore, for the proposed conditions $P_{2A}$ should be equal or lower than 1.976 atm.

According to Ec. (16), the value of $P_{2A}$, also depends on the variation of the total entropy of the universe. For example, if we use a trajectory with $dSu = 0.187$ cal/K, we get $P_{2A} = 1.8$ atm. But, if we use a trajectory with $dSu = 0.998$ cal/K, then $P_{2A} = 1.2$ atm. In any event, since the process is irreversible, $P_{2A}$ should be lower than 1.976 atm in order to occur according to the global formulation of the second law of thermodynamics.

An interesting aspect of the process illustrated in Fig. 1 is the work $W_A$ obtained from the isothermal expansion of tank A. Is it convenient to indicate that the external pressure Pext of tank A is not known, then we are not able to calculate the work $W_A$ using the conventional expression used for irreversible processes $W_A = \int P_{ext} dV$, where V is the volume of the system. However, we can estimate it by applying the first law of thermodynamics to the processes that happen in tanks A and B. Thus

$$W_A = Q_A = -Q_B = -Cv (T_{2B} - T_{1B}) \tag{18}$$

where $Q_A$ and $Q_B$ represent the heat transferred among tanks A and B.

In accordance with Ec. (18) the work $W_A$ produced by the isothermal expansion of tank A is 900 cal. We may detect that for the trajectories that lead to pressures $P_{2A}$ greater than 1.626 atm and lower than 1.976 atm, the work produced by the process is greater than the obtained for a reversible expansion involving the same change of state. For example, for a trajectory allowed by the global formulation of the second law of thermodynamic that leads to $P_{2A} = 1.8$ atm, the irreversible process illustrated in Fig.1 does a work of 900 cal. While the work made by a reversible expansion among the same change of state is 798.5 cal. This is an unusual result from the point of view of classical thermodynamics, but it can be explained, by arguing an entropy coupling between the process taking place in tanks A and B. This argument will be presented in the next paragraphs. On the other hand, for trajectories that conduct to pressures $P_{2A}$ lower than 1.626 atm, the results agree with the conventional expectations of classical thermodynamics.

To explain the previous results, we can apply Ec. (14) to the proposed universe, obtaining

$$d_iSu = d_iSm + d_iS_A + d_iS_B \tag{19}$$

Here, $d_i S_A$ and $d_i S_B$ symbolize the production of internal entropy in tanks A and B, respectively, and $d_i Sm$ is the entropy production due to the heat flow between the tanks A and B.

By substituting, combining and integrating Ecs. (1), (10), (11), (13), (14) and (19), we get

$$\Delta_i S_A = - R \ Ln \ (P_{2A} / P_{1A}) - Q_A / T_A \quad (20)$$

$$\Delta_i S_B = 0 \quad (21)$$

$$\Delta_i Sm = Q_A / T_A + Cv \ Ln \ (T_{2B} / T_{1B}) \quad (22)$$

$$\Delta_i Su = - R \ Ln \ (P2_A / P_{1A}) + Cv \ Ln \ (T_{2B} / T_{1B}) \quad (23)$$

Here, $\Delta_i S_A$, $\Delta_i S_B$, $\Delta_i Sm$ y $\Delta_i Su$ are the internal entropy production during the specified change of state for tank A, tank B, for the heat flow between both tanks and for the universe of the process, respectively.

As we may detect Ecs. (16) and (23) coincide, just as Ec (8) suggests for the case of an adiabatic closed universe.                                                                                                      .

By combining Ecs. (18) and (20) we can relate the work $W_A$ with the production of internal entropy in tank A by means of the following equation

$$W_A = - T_A \ \Delta_i S_A - R \ T_A \ Ln \ (P_{2A} / P_{1A}) \quad (24)$$

We can notice that for pressures lower than 1.626 atm, $\Delta_i S_B$ is equal to zero, but $\Delta_i Sm$, $\Delta_i S_A$ y $\Delta_i Su$ are quantities greater than zero, and the work $W_A$ done by the proposed irreversible transformation is lower than the reversible work Wr for the same change of state. For example, for a trajectory that leads to a value of $P_{2A} = 1.2$ atm, we find $\Delta_i Sm = 0.390$ cal/K, $\Delta_i S_A = 0.608$ cal/K, $\Delta_i S_B = 0$ cal/k, $\Delta_i Su = 0.998$ cal/K, $W_A = 900$cal and Wr =1204 cal. The conventional thermodynamics explains very well this behavior, arguing that when internal entropy is produced in an irreversible process, this loses capacity to produce work in comparison with a reversible operation under the same change of state. The difference between the reversible work Wr and the irreversible work $W_A$ is known as the lost work Wp, and in the case of an isothermal transformation in a closed system it can be demonstrated that Wp = $T_A \ \Delta_i S_A$. Under these considerations Ec. (24) is equivalent to

$$W_A = - Wp + Wr \quad (25)$$

In the proposed example Wp =304 cal., Wr = 1204 cal and $W_A = 900$cal

On the other hand, we observe that for trajectories allowed by the global formulation that reach pressures $P_{2A}$ greater than 1.626 atm and lower than 1.978 atm, $\Delta_i Sm$ and $\Delta_i Su$ are quantities greater than zero, $\Delta_i S_B$ is equal to zero, but $\Delta_i S_A$ is lower than zero, and the work $W_A$ executed by the proposed irreversible transformation is greater than the reversible work Wr for the same change of state. By inspecting, it is observed that for the case of a specific pressure inside the indicated range as $P_{2A}$ = 1.8 atm, we find that $\Delta_i S_m$ = 0.390 cal/K, $\Delta_i S_A$ = - 0.203 cal/K, $\Delta_i S_B$ = 0 cal/k , $\Delta_i S_u$ = 0.187 cal/K, $W_A$ = 900cal and Wr =798.5 cal. Since the entropy production in tank A is negative we can argue, analogously to the previous case, that the entropy destruction allows to the system to win an additional work Wg = - $T_A \Delta_i S_A$, in comparison with the reversible work Wr, and Ec. (24) becomes

$W_A$ = Wg + Wr  (26)

By examining, it is observed that for the case $P_{2A}$ = 1.8 atm, Wr = 798.5 cal, Wg = 101.5 cal and $W_A$ = 900 cal.

It is also detected that the process can reach a stationary state in which the negative production of entropy is equal to the positive production of entropy, but different from zero. In these circumstances $\Delta_i S_A + \Delta_i S_B + \Delta_i S_m$ = 0, and $\Delta S_u = \Delta_i S_u$ = 0. This state is reached when $\Delta_i S_m$ = 0.390 cal/K, $\Delta_i S_A$ = - 0.390 cal/K, and $\Delta_i S_B$ = 0 cal/k. The trajectory of this stationary state leads to a final pressure $P_{2A}$ = 1.976 atm. Under this trajectory the work done by the irreversible isothermal expansion in tank A is $W_A$ = 900 cal. The corresponding reversible work under the same change of state is Wr = 705 cal and the work won is Wg = 195 cal. As we can see, this trajectory is, thermodynamically, the most efficient route we can find for the proposed process, and the final state achieved corresponds to a dynamic stationary state in which the positive entropy production is compensated, exactly, by the negative entropy production. In this condition, the universe remains at constant entropy operating irreversibly under finite gradients of the thermodynamic variables.

In general, following similar procedures, it is possible to design different versions of entropy couplings in closed and open systems operating under isobaric, isochoric, isothermal and adiabatic conditions, among other permissible alternatives. Some of these versions have been published previously [4, 5].

5. Conclusions

According to the propositions of this paper, the combination of the local and global formulations of the second law of thermodynamics suggests the possibility of theoretical existence of irreversible processes with entropy couplings among the different parts of the universe. Such transitions allowed by the combined formulation of the second law of thermodynamics produce unexpected effects from the point of view of conventional thermodynamics as the possibility of being more efficient than a reversible operation under the same change of state. The maximum efficiency of these transformations is obtained when the positive internal entropy production compensates the negative entropy production reaching a stationary state unpredicted by classical thermodynamics.

It is convenient to indicate that when the local or the global formulations of the second law of thermodynamics are applied in an independent way it is not possible to predict the entropy couplings analyzed here. Only, a combination of both formulations in the sense proposed in this study suggests this interesting possibility.

References

[1] Prigogine I., Thermodynamic of Irreversible Processes, 3 rd edition, Interscience Publishers, New York,1967.

[2] Smith J.M. and Van Ness H.C., Introduction to Chemical Engineering Thermodynamics, 3 rd edition, McGraw Hill Book Company, New York, 1975.

[3] Van Wylen G.J. and Sonntag R. E., Fundamentals of Classical Thermodynamics, 1st edition, John Wiley and Sons, New York, 1965.

[4] Belandria J.I., J.Chem. Educ., 72, 116,1995

[5] Belandria J.I., Europhys.Lett., 70(4), 446-451 , 2005